\documentclass[12pt,a4paper]{article}

\usepackage[margin=2.5cm]{geometry}
\usepackage{amsmath,amssymb,amsthm}
\usepackage{mathtools}
\usepackage{hyperref}
\hypersetup{colorlinks=true, linkcolor=blue, citecolor=blue, urlcolor=blue}
\usepackage{microtype}
\usepackage{xcolor}
\usepackage{titlesec}
\usepackage{cite}

\titleformat{\section}{\large\bfseries}{\thesection.}{0.5em}{}
\titleformat{\subsection}{\normalsize\bfseries}{\thesubsection}{0.5em}{}

\newcommand{\ket}[1]{\left|#1\right\rangle}
\newcommand{\comm}[2]{\left[#1,\,#2\right]}
\newcommand{\acomm}[2]{\left\{#1,\,#2\right\}}
\newcommand{\half}{\tfrac{1}{2}}

\begin{document}

%% ---- Title block ----------------------------------------------------

\vspace{1em}

\begin{center}
  {\LARGE\bfseries
 Absence of Continuous Spin Particles in Superstring Theory}

  \vspace{1.8em}

  {\bf Arwa Alabbasi$^{1}$, Fernando Quevedo$^{1,2}$}

  \vspace{0.8em}

  {\small
    $^{1}$\textit{Department of Physics, New York University Abu Dhabi, Abu Dhabi, United Arab Emirates}\\
    $^2$ \textit{DAMTP, University of Cambridge, Wilberforce Road, Cambridge, CB3 0WA, UK}
%    \texttt{aaa9822@NYU.edu}
  }

  \vspace{1.5em}

  \textbf{Abstract}
\end{center}

\begin{quote}\small
The study of continuous spin particles, as general low-energy implications of relativity and  quantum mechanics, has intensified in recent years  with potential experimental signatures. In this note we show that these particles are absent in perturbative superstring theories, generalising a result in \cite{Font:2013}  to the supersymmetric case. This stands out as a general low-energy prediction of all perturbative superstring constructions.

\end{quote}

\vspace{1em}
\hrule
\vspace{1.5em}

\tableofcontents 
%% =====================================================================
\section{Introduction}
\label{sec:intro}
%% =====================================================================
The most general structure of massless states implied by $4d$ Poincar\'e invariance and quantum mechanics depends on a continuous label $\rho$ corresponding to the eigenvalue of the second Casimir operator (the first label being the mass). Only for $\rho=0$ we recover the standard spectrum of massless particles classified by their momentum $p^\mu$ and helicity $\lambda$. For the generic case $\rho\neq 0$ the corresponding particles are historically (and somewhat misleadingly) called continuous spin particles (CSP) \cite{Wigner:1939,Weinberg:1995}. This raises a fundamental theoretical question: Do these particles exist in nature, and if not, why? 

Over the past decade, a series of papers by Schuster, Toro and collaborators, has vastly advanced our understanding of CSPs interactions and their potential observational signatures \cite{Schuster:2013pxj, Schuster:2013vpr,Schuster:2013pta, Schuster:2014hca, Schuster:2023xqa, Schuster:2023jgc, Schuster:2024wjc, Kundu:2025fsd, Reilly:2025lnm, Kundu:2025mzm}. In particular, they have demonstrated the following:
\begin{itemize}
\item Despite comprising an infinite tower of states of different helicities, CSP interactions can be described by some local actions \cite{Schuster:2013pta, Schuster:2014hca}. Consistent couplings to matter have also been constructed in flat spacetime \cite{Bekaert:2017khg}.

\item  Lorentz invariant, tree-level scattering amplitudes for CSP emissions and exchange can be constructed \cite{Schuster:2013pxj, Schuster:2013vpr}.
\item 
At high energies ($E\gg \rho$) CSP scattering and radiation amplitudes approach those of standard helicity $\lambda=0,1,2$ massless particles \cite{Schuster:2013vpr,Schuster:2014hca, Schuster:2023xqa}.  
 The additional helicity modes act as weakly coupled dark sector, naturally  explaining why CSPs remain hidden while still producing standard long-range interactions. 
\item They provided a counter argument to Wigner's original objection   regarding  infinite heat capacity of the vacuum showing that at weak couplings only finite heat capacity is needed to excite these states \cite{Schuster:2024wjc}.

\item The extra helicity modes modify QED opening new channels for atomic transitions \cite{Reilly:2025lnm} and introducing corrections to Larmor's formula \cite{Schuster:2023xqa} for synchrotron radiation that can be tested experimentally. Similarly at energy scales of order $\rho$ they find testable corrections to soft photon or graviton Bremsstrahlung. 

\item They exhibited a candidate CSP–graviton matrix element \cite{Schuster:2013vpr} demonstrating that the Weinberg–Witten theorem, which restricts how massless higher-helicity particles can couple to gravity, does not apply to CSPs, removing a long-standing theoretical obstruction.

\item 
The program has since been extended beyond scalar matter \cite{Schuster:2023jgc,Kundu:2025mzm}: explicit, unitarity-checked scattering rules for CSP-photon QED (pair annihilation, Compton scattering) have been derived, and a supersymmetric worldline formalism now couples continuous-spin fields to spin $\frac12$ matter, reducing to ordinary QED or a Yukawa interaction in the appropriate limits.

\item 
More recently \cite{Kundu:2025fsd}, the framework has been extended to gravity itself: coupling matter to a continuous-spin graviton at linearized level predicts a time-delay signature of order $\rho_g/\omega$ in gravitational wave interferometers (where $\rho_g$ is the CSP parameter for the graviton and $\omega$ the gravitational wave frequency), offering a short-term observational test.

\end{itemize}

Further work in parallel by Metsaev has included couplings, supersymmetry and extensions to AdS (see for instance: \cite{Metsaev:2016lhs, Metsaev:2017cuz, Metsaev:2025nbm, Metsaev:2025wcv}). Related developments by Bekaert, Najafizadeh and collaborators include a covariant action for fermionic CSPs, connections between the Schuster--Toro and Metsaev formulations, and both on-shell and off-shell supersymmetric extensions of continuous-spin gauge theories \cite{Bekaert:2015tva,Najafizadeh:2017tin,Najafizadeh:2019mun,Najafizadeh:2021tjc}

More recently, on-shell scattering amplitudes for CSPs have also been studied in \cite{Bellazzini:2024dco}
and a spinor-helicty formalism was introduced in \cite{Bellazzini:2024dco, Basile:2026ntj} to systematically study CSP interactions.
These are remarkable achievements that indicate that CSPs represent a viable, highly structured, and testable infrared (IR) extension of the Standard Model.

On the ultraviolet (UV) side,  string theory  provides a consistent quantum completion of field theories including gravity. At low energies, the massless spectrum of string theories, after compactification to four dimensions, precisely gives single particle states  with helicities $\lambda=0,\frac12,1,\frac{3}{2},2$ explaining from a single stroke the structure derived in field theory using different tools, such as gauge symmetries and soft-theorems, and corresponding with the particles observed so far (except for $\frac{3}{2}$ and strictly speaking the graviton has not explicitly been discovered as a quantum object). This is one of the remarkable triumphs of string theory \footnote{In the massive case, the existence of an infinite tower of massive modes is another implication from string theory that fits with the efforts to make sense  of UV completion of massive higher spin particles in which an  infinite tower seems to be required for consistency (see for instance \cite{Duff:1989ea,Huang:2022mdb}).}

It is therefore natural to inquire if CSPs can be obtained from string theory. In a previous publication \cite{Font:2013} it was demonstrated that CSPs are strictly absent in perturbative bosonic string theory. That is bosonic strings uniquely select $\rho=0$. It is interesting that following different techniques, the analysis of \cite{Bellazzini:2024dco} is consistent with those findings.  In this note we extend this result to the supersymmetric regime, establishing that CSPs are not present in all formulations of string theory so far (with the possible exception of  tension-less strings). Therefore, modulo unknown potential non-perturbative effects, superstring theory universally predicts the absence of CSPs.

This result carries significant implications:
\begin{itemize}
\item {\it Falsifiability}. String theory famously has a shortage of general predictions\footnote{Only a handful may  count as generic predictions such as the existence of gravity, gauge interactions, moduli fields, no more than 11-dimensions and supersymmetry at an unkown scale.} that can rule-out the framework. Any positive experimental signal  of CSPs would directly falsify perturbative string theory.
\item {\it Theoretical selection}. If CSPs are absent in nature, string theory offers the explanation of why they do not exist, fulfilling the main goal of theoretical physics of explaining experimental results with the minimum number of assumptions.
\item {\it Elevated experimental motivation}. This finding heightens the stakes for the CSP searches proposed by  Schuster, Toro and collaborators. Experimental tests of CSPs acquire extra relevance  as also potential direct  tests of string theory itself. 
\end{itemize}
Notice that the generality of this result bypasses the swampland  programme \cite{Vafa:2005ui} since, despite the progress mentioned above, the existence of a full low energy effective field theory (EFT) for CSPs is not yet available and the swampland programme concerns with the consistency of EFTs without UV completion.
Our conclusions do not rely on EFT consistency conditions, but rather follow directly from the UV dynamics of the string worldsheet.

In the next section we briefly summarise the origin of continuous spin particles in arbitrary dimensions $d$ for which the Little Group is the inhomogenous orthogonal group $ISO(d-2)$ comprising a semi-direct product of the standard rotations in $SO(d-2)$ and translations. On section 3 we discuss NSR supersymmetric string theory formalism in the light-cone gauge, identifying the relevant rotation $S_{ij}$ and translation $T_i$ operators in terms of the string bosonic and fermionic string oscillators. Sections 4 and 5 show how the translation operators $T_i$ annihilate the massless states for the massless NS and R states respectively.
Section 6 extends the results to closed superstrings, IIA, IIB and heterotic.

In both cases we will see that $T^i = 0$ acting on massless states implies
that  the little group collapses from
$ISO(8)$ to $SO(8)$, confirming the absence of CSPs from the superstring
spectrum. This introduces a falsifiable, compactification-independent
prediction of the theory so should particles carrying continuous spin be detected
experimentally we would be able to  rule out perturbative superstring constructions.

%% =====================================================================
\section{Review of Continuous Spin Representations}
\label{sec:review}
%% =====================================================================

The generators of the
Poincar\'e group satisfy
\begin{align}
  \comm{M^{\mu\nu}}{M^{\alpha\beta}} &=
    i\bigl(\eta^{\mu\alpha}M^{\nu\beta} - \eta^{\nu\alpha}M^{\mu\beta}
          + \eta^{\nu\beta}M^{\mu\alpha} - \eta^{\mu\beta}M^{\nu\alpha}\bigr),\\
  \comm{M^{\mu\nu}}{P^\alpha} &=
    i\bigl(\eta^{\mu\alpha}P^\nu - \eta^{\nu\alpha}P^\mu\bigr),
  \qquad \comm{P^\alpha}{P^\beta} = 0,
  \label{eq:Poincare1}
\end{align}
where Greek indices run from $0$ to $d-1$. In the light-cone frame,
\begin{equation}
  P^\pm = \frac{1}{\sqrt{2}}\bigl(P^0 \pm P^{d-1}\bigr), \qquad
  M^{\pm i} = \frac{1}{\sqrt{2}}\bigl(M^{0i} \pm M^{(d-1)i}\bigr),
  \quad i = 1,\ldots,d-2.
\end{equation}
For massless particles the reference momentum is $p^\mu = (E,0,\ldots,0,E)$,
so $p^+ = \sqrt{2}E$ and $p^- = p^i = 0$ and the generators stabilizing this
momentum are $M^{ij}$ and $\Pi^i \equiv M^{-i}$, satisfying the $ISO(d-2)$
algebra
\begin{equation}
  \comm{M^{ij}}{M^{kl}} =
    i\bigl(\delta^{ik}M^{jl} - \delta^{jk}M^{il}
          + \delta^{jl}M^{ik} - \delta^{il}M^{jk}\bigr),
  \label{eq:ISOrot}
\end{equation}
\begin{equation}
  \comm{M^{ij}}{\Pi^k} =
    i\bigl(\delta^{ik}\Pi^j - \delta^{jk}\Pi^i\bigr),
  \qquad
  \comm{\Pi^i}{\Pi^j} = 0.
  \label{eq:ISOfull}
\end{equation}
States in the helicity representation satisfy $\Pi^i|\psi\rangle = 0$,
reducing the little group to $SO(d-2)$. States in a CSR satisfy
$\sum_i(\Pi^i)^2 \neq 0$ and carry an additional continuous label
$\theta \in [0,2\pi)$ \cite{Font:2013}.

For  a supersymmetric relativistic particle in the light-cone gauge the generator $\Pi^i$ decomposes as \cite{Brink:2002zx}
\begin{equation}
  \Pi^i = x^- p^i - \tfrac{1}{2}\acomm{x^i}{p^-}
            + \frac{1}{p^+}\bigl(T^i - p^j S^{ij}\bigr)
+i\,\frac{p^i}{2p^+}
\left(
\theta\frac{\partial}{\partial\theta}
+
\bar{\theta}\frac{\partial}{\partial\bar{\theta}}
\right)
,
  \label{eq:Mminus}
\end{equation} 
where $\bar{\theta}$ is the complex conjugate of the anticommuting Grassmann parameter $\theta$, $S^{ij}$ $SO(d-2)$ spin generators and the operators $T^i$
satisfy
\begin{equation}
  \comm{S^{ij}}{T^k} = i\bigl(\delta^{ik}T^j - \delta^{jk}T^i\bigr),
  \qquad
  \comm{T^i}{T^j} = im^2 S^{ij}
  \label{eq:Talgebra}
\end{equation} When $m=0$ the $T^i$ commute and become the translation generators of
$ISO(d-2)$. \\

We notice that in the chosen frame ($p^i=0$), all the \(\Pi^i\) terms go to zero trivially except \(\frac{T^i}{p^+}\)\footnote{To establish $\left\{x^i,p^-\right\}|\psi\rangle=0$ it requires the use of $\left\{ x^k, (p^k)^2 \right\}=2x^k(p^k)^2-2ip^k$.}.
Therefore showing that $T^i|\psi\rangle = 0$ for massless string states
is sufficient to demonstrate the triviality of the action of $ \Pi^i$ and therefore absence of CSPs in string theory that we discuss next.

%\paragraph{Summary of results.}

%% =====================================================================
\section{The NSR Superstring in Light-Cone Gauge}
\label{sec:superstring}
%% =====================================================================
In this and the next section, working in the light-cone gauge NSR formalism in $d = 10$ superstring theory, we will  prove:
\begin{enumerate}
  \item \textbf{NS sector:} $T^i = T^i_B + T^i_F$ annihilates the massless
        vector state $|j\rangle_{\rm NS} = \psi^j_{-1/2}|p^+,\vec{p}\,\rangle$, or $|j\rangle_{\rm NS} = \psi^j_{-1/2}|0\rangle$
        with both contributions vanishing separately in the frame $p^i = 0$.
  \item \textbf{R sector:} $T^i$ annihilates the massless Ramond ground
        state $|A\rangle_R$, again with both contributions vanishing
        independently.
\end{enumerate}
We will first discuss the open string and will comment on the closed string at the end.

In light-cone gauge $X^+ = \tau$, $p^+ = \text{const}$. The transverse fields $X^i$ and
$\psi^i$ ($i = 1,\ldots,8$) are expanded as
\begin{equation}
  X^i(\sigma,\tau) \;\sim\; x^i_0 + 2\alpha' p^i \tau
  + i\sqrt{\tfrac{\alpha'}{2}}\sum_{n\neq 0}\frac{\alpha^i_n}{n}
    e^{-in\tau}\cos(n\sigma),
  \label{eq:Xmode}
\end{equation}
with $\comm{\alpha^i_m}{\alpha^j_{-n}} = m\delta^{ij}\delta_{m,n}$,
$\comm{x^i_0}{p^j} = i\delta^{ij}$, and $\comm{x^-_0}{p^+} = -i$. The mode expansion is understood here for NN boundary conditions; for DD conditions the oscillator basis is instead \(\sin(n\sigma)\), while mixed ND and DN conditions yield half-integer moding, with \(\cos(r\sigma)\) and \(\sin(r\sigma)\), respectively, \(r\in\mathbb{Z}+\tfrac12\).

\paragraph{NS sector.}
Fermionic modes are half-integer moded,
\begin{equation}
  \psi^i(\sigma,\tau) \;\sim\;
    \sum_{r\in\mathbb{Z}+\frac{1}{2}} \psi^i_r\, e^{ir\tau}\cos(r\sigma),
  \qquad
  \acomm{\psi^i_r}{\psi^j_s} = \delta^{ij}\delta_{r+s,0}.
  \label{eq:NSmode}
\end{equation}
The mass-shell condition is $\alpha' M^2 = N_B + N_F - \half$ with
$N_B = \sum_{n>0}\alpha^i_{-n}\alpha^i_n$ and
$N_F = \sum_{r>0} r\, \psi^i_{-r}\psi^i_r$.
The massless states at $N_B + N_F = \half$ are
\begin{equation}
  |j\rangle_{\rm NS} \;=\; \psi^j_{-1/2}|p^+,\vec{p}\,\rangle,
  \label{eq:NSmassless}
\end{equation}
transforming as a vector of $SO(8)$.

\paragraph{R sector.}
Fermionic modes are integer-moded, $r\in\mathbb{Z}$, so zero modes $\psi^i_0$
exist and satisfy $\acomm{\psi^i_0}{\psi^j_0} = \delta^{ij}$. The Ramond ground
state $|A\rangle_R$ is defined by
\begin{equation}
  \alpha^i_n|A\rangle_R = 0\;\;(n>0), \qquad
  \psi^i_n|A\rangle_R = 0\;\;(n>0),
  \label{eq:Rvac}
\end{equation}
and transforms as an $SO(8)$ spinor under $\psi^i_0$. The normal-ordering
constant vanishes in the R sector, so $|A\rangle_R$ is massless.

\subsection{Lorentz Generators and Fermionic contribution to \(T^i\)}

To see where the fermionic part of the light-cone Lorentz generator comes from, we start from the standard NSR expression \cite{GreenSchwarzWitten1987}\footnote{See section 4.3, ``Light-cone gauge quantization,'' pp.~212--214,
especially eqs.~(4.3.9), (4.3.14), (4.3.15), and (4.3.16).}. Decomposing the generators $M^{\mu\nu}$ as:
\begin{equation}
M^{\mu\nu}=l^{\mu\nu}+E^{\mu\nu}+ K^{\mu\nu}
\end{equation}
with $l^{\mu\nu}$ the orbital component, $E^{\mu\nu}$ the bosonic contribution and $K^{\mu\nu}$ the fermionic contribution, we can write:
\begin{equation}
    K^{i-}
    =
    \frac{1}{p^+}
    \sum_{n=-\infty}^{\infty}
    K^{ij}_{-n}\alpha^j_n ,
\end{equation} 
where
\begin{equation}
    K^{ij}_m
    =
    -\frac{i}{2}
    \sum_r
    \left(
        \psi^i_{m-r}\psi^j_r
        -
        \psi^j_{m-r}\psi^i_r
    \right).
\end{equation}
Here \(r\in\mathbb{Z}+\frac12\) in the NS sector and \(r\in\mathbb{Z}\) in the R sector.

Using $ \tilde\psi_r^{-}=\frac{1}{p_+}\sum \alpha^i_{r-s}\psi^i_s$ \cite{GreenSchwarzWitten1987}, the above expression may be rewritten as
\begin{equation}
    K^{i-}
    =
    -i
    \sum_{r>0}
    \left(
        \psi^i_{-r}\tilde\psi^-_r
        -
        \tilde\psi^-_{-r}\psi^i_r
    \right),
\end{equation}
up to the overall sign convention for the Lorentz generators. Then, to identify the corresponding contribution to \(T^i\), we compare two
decompositions of the same light-cone Lorentz generator $M^{i-}$. On the one hand,
the little-group decomposition gives \eqref{eq:Mminus}.  
Thus the oscillator-spin part of \(M^{-i}\) is
\begin{equation}
    \frac{1}{p^+}
    \left(
        T^i-p^jS^{ij}
    \right).
\end{equation}
On the other hand, in the light-cone NSR string, the same generator is
written as
\begin{equation}
    M^{i-}=l^{i-}+E^{i-}+K^{i-},
\end{equation}
%where \(l^{i-}\) is the orbital zero-mode contribution, \(E^{i-}\) is the
%bosonic oscillator contribution, and \(K^{i-}\) is the fermionic oscillator
%contribution. 
Since the orbital pieces match each other, the remaining
oscillator-spin pieces must satisfy
\begin{equation}
    \frac{1}{p^+}
    \left(
        T^i-p^jS^{ij}
    \right)
    =
    E^{i-}+K^{i-}
\label {17}\end{equation}
Splitting \(T^i=T_B^i+T_F^i\) and \(S^{ij}=S_B^{ij}+S_F^{ij}\), the
fermionic part is therefore

\begin{equation}
    T_F^i
    =
    p^+K^{i-}
    +
    p^jS_F^{ij}=i
    \sum_{r>0}
    \left[
        \psi^i_{-r}
        \left(
            p^+\tilde\psi^-_r-p^j\psi^j_r
        \right)
        -
        \left(
            p^+\tilde\psi^-_{-r}-p^j\psi^j_{-r}
        \right)
        \psi^i_r
    \right],
    \label{eq:TF-final}
\end{equation} In the massless little-group frame \(p^i=0\), this reduces to
\begin{equation}
    T_F^i
    =
    ip^+
    \sum_{r>0}
    \left(
        \psi^i_{-r}\tilde\psi^-_r
        -
       \tilde\psi^-_{-r}\psi^i_r
    \right).
\end{equation}
Summarising, the $T^i$ operator splits as $T^i = T^i_B + T^i_F$ \cite{Font:2013} , with
\begin{align}
  T^i_B &= i\sum_{n=1}^\infty\frac{1}{n}
    \Bigl[\alpha^i_{-n}\bigl(p^+\alpha^-_n - p^j\alpha^j_n\bigr)
         -\bigl(p^+\alpha^-_{-n} - p^j\alpha^j_{-n}\bigr)\alpha^i_n
    \Bigr] ,\\
  T^i_F &= i\sum_{r>0}
    \Bigl[\psi^i_{-r}\bigl(p^+\tilde\psi^-_r - p^j \psi^j_r\bigr)
         -\bigl(p^+\tilde\psi^-_{-r} - p^j \psi^j_{-r}\bigr)\psi^i_r
    \Bigr],
  \label{eq:TiF}
\end{align}
where
\begin{equation}
  \tilde\psi^-_r = \frac{1}{p^+}
    \sum_{s} \alpha^k_{r-s}\,\psi^k_s.
  \label{eq:psitilde}
\end{equation}
And the sum over $s$ runs over half-integers in the NS sector and integers in
the R sector.

%% =====================================================================
%\include{NSR}

%% ============================================================
\section{NS Sector: $T^i$ on the Massless Vector State}
\label{sec:NS}
%% =====================================================================

We work in the massless little-group frame $p^i = 0$ throughout, which
simplifies $T^i_B$ and $T^i_F$ to terms involving $p^+\alpha^-_n$ and
$p^+\tilde\psi^-_r$ only.

\subsection{Bosonic contribution $T^i_B$}

We evaluate the bosonic oscillator contribution
\begin{equation}
T_B^i
=
i\sum_{n=1}^{\infty}\frac{1}{n}
\left[
\alpha^i_{-n}p^+\alpha^-_n
-
p^+\alpha^-_{-n}\alpha^i_n
\right],
\end{equation}
Although this is the bosonic contribution to the Lorentz
generator, the light-cone constraint expresses $\alpha^-_n$ in terms of
both transverse bosonic and fermionic oscillators.  In the NS sector,
\begin{equation}
p^+\alpha^-_n
=
\frac{1}{2}
\sum_{m=-\infty}^{\infty}
\alpha^k_{n-m}\alpha^k_m
+
\frac{1}{2}
\sum_{r\in \mathbb{Z}+\frac12}
\left(r-\frac{n}{2}\right)
\psi^k_{n-r}\psi^k_r ,
\label{eq:alpha-minus-NS}
\end{equation}
Therefore,
\begin{equation}
T_B^i |j\rangle_{\rm NS}
=
i\sum_{n=1}^{\infty}\frac{1}{n}
\left[
\underbrace{\alpha^i_{-n}p^+\alpha^-_n
\psi^j_{-\frac12}|0\rangle}_{(I)}
-
\underbrace{p^+\alpha^-_{-n}\alpha^i_n
\psi^j_{-\frac12}|0\rangle}_{(II)}
\right].
\end{equation}

First, let us consider
\begin{equation}
(I)
=
\alpha^i_{-n}p^+\alpha^-_n
\psi^j_{-\frac12}|0\rangle .
\end{equation}
Using \eqref{eq:alpha-minus-NS}, this becomes

\begin{equation}
(I)
=
\frac{1}{2}\alpha^i_{-n}
\sum_m
\alpha^k_{n-m}\alpha^k_m
\psi^j_{-\frac12}|0\rangle
+
\frac{1}{2}\alpha^i_{-n}
\sum_{r\in\mathbb{Z}+\frac12}
\left(r-\frac{n}{2}\right)
\psi^k_{n-r}\psi^k_r
\psi^j_{-\frac12}|0\rangle .
\end{equation}

The first term here vanishes because for $n>0$,  $m<0$ means the first bosonic oscillator is an annihilator that can commute forward and act directly on the vacuum and $m>0$ means the second bosonic oscillator is an annihilator. \\

We can show the same for the second term. For the same reason, between $\psi^k_{n-r}$ and  $\psi^k_{r}$ there is always at least one annihilator which can then commute to act directly on the vacuum, the only case in which it does not commute is $r=\frac12$:

\begin{equation}
\left(\frac{1-n}{2}\right)
\psi^k_{n-\frac12}\psi^k_{\frac12}
\psi^j_{-\frac12}|0\rangle
=\left(\frac{1-n}{2}\right)
\psi^k_{n-\frac12}[\eta^{jk}\delta_{\frac12,\frac12}-\psi^j_{-\frac12}\psi^k_{\frac12}]=\left(\frac{1-n}{2}\right)
\psi^k_{n-\frac12}[\eta^{jk}-\psi^j_{-\frac12}\psi^k_{\frac12}]
\end{equation}

we know that 
\begin{equation}
    \{\psi_r^\mu,\psi_s^\nu\} = \eta^{\mu\nu}\delta_{r,-s}
    \quad\Rightarrow\quad
    \{\psi_{-\frac{1}{2}}^j,\psi_s^k\} = \eta^{jk}\delta_{-\frac{1}{2},-s}
    \quad\Rightarrow\quad
    \psi_s^i\psi_{-\frac{1}{2}}^j = \eta^{ij}\delta_{s,\frac{1}{2}} - \psi_{-\frac{1}{2}}^j\psi_s^i
    \label{eq:commutator}
\end{equation}

The remaining fermionic oscillator is
%\begin{equation}
$\psi^k_{n-\frac12}$.
%\end{equation}
Since $n\geq 1$, we have $n-\frac12>0$, so this is an annihilation
operator and kills the vacuum. Therefore the $r=\frac12$ term also
vanishes.

For all other values of $r$, no contraction removes the annihilation
operator structure. It follows that $(I)=0$.

Now consider the second term,
\begin{equation}
(II)
=
p^+\alpha^-_{-n}\alpha^i_n
\psi^j_{-\frac12}|0\rangle .
\end{equation}
The transverse bosonic oscillator $\alpha^i_n$ commutes with the fermionic
oscillator $\psi^j_{-\frac12}$, and for $n>0$ it annihilates the vacuum, therefore
%\begin{equation}
$(II)=0$.
%\end{equation}

Since both terms vanish for every positive integer $n$, we conclude that
\begin{equation}
T_B^i |j\rangle_{\rm NS}=0 .
\end{equation}
Thus the bosonic oscillator contribution to the null-rotation generator
annihilates the NS massless vector state.

\subsection{Fermionic contribution $T^i_F$}

Setting $p^i = 0$ in \eqref{eq:TiF} and focusing on the lowest mode
$r = \half$ (lowest positive energy mode, indeed the massless case), we obtain
\begin{equation}
  T^i_F|j\rangle_{\rm NS}
  = i\Bigl[
      \underbrace{\psi^i_{-1/2}\,p^+\tilde\psi^-_{1/2}\,\psi^j_{-1/2}|0\rangle}_{(I)}
     -\underbrace{p^+\tilde\psi^-_{-1/2}\,\psi^i_{1/2}\,\psi^j_{-1/2}|0\rangle}_{(II)}
    \Bigr].
  \label{eq:split}
\end{equation}

\paragraph{Term (I).}
Expanding $\tilde\psi^-_{1/2}$,
\begin{equation}
  (I) = \frac{(p^+)^2}{\sqrt{2\alpha'}}
    \,\psi^i_{-\frac12}\sum_{s\in\mathbb{Z}+\frac{1}{2}}
    \alpha^k_{\frac12-s}\,\psi^k_s\,\psi^j_{-\frac12}|0\rangle.
\end{equation}
and using \eqref{eq:commutator}: 
\begin{equation}
  (I) = \frac{(p^+)^2}{\sqrt{2\alpha'}}
    \,\psi^i_{-\frac1
    2}\sum_{s\in\mathbb{Z}+\frac{1}{2}}
    \alpha^k_{\frac12-s}\,[\eta^{jk}\delta_{\frac12,s}-\psi^k_{-\frac12}\psi^j_s]|0\rangle.
\end{equation}

\[
\text{explicitly }:
\qquad
\begin{cases}
s=\tfrac12:
&
\psi^i_{-\frac12}
    \alpha^k_{\frac12-\frac12}\,[\eta^{jk}\delta_{\frac12,\frac12}-\psi^k_{-\frac12}\psi^j_\frac12]|0\rangle=
    \psi^i_{-\frac12}
    \alpha^k_0\,[\eta^{jk}-\psi^k_{-\frac12}\psi^j_\frac12]|0\rangle=0
\\[0.8ex]
&
\\[2.2ex]
s> \frac12:
&
-\psi^i_{-\frac1
    2}
    \alpha^k_{\frac12-s}\,[\eta^{jk}\delta_{\frac12,s}-\psi^k_{-\frac12}\psi^j_s]|0\rangle=\psi^i_{-\frac1
    2}
    \alpha^k_{\frac12-s}\,\psi^k_{-\frac12}\psi^j_s|0\rangle=0
\\[0.8ex]
&
\\[2.2ex]
s<0:
&
-\psi^i_{-\frac1
    2}
    \alpha^k_{\frac12-s}\,[\eta^{jk}\delta_{\frac12,s}-\psi^k_{-\frac12}\psi^j_s]|0\rangle=\psi^i_{-\frac1
    2}
    \alpha^k_{\frac12-s}\,\psi^k_{-\frac12}\psi^j_s|0\rangle
    \\[0.8ex]&
    =\psi^i_{-\frac1
    2}\psi^k_{-\frac12}\psi^j_s\alpha^k_{\frac12-s}|0\rangle=0
\end{cases}
\]

\[
\Rightarrow (I)=0
\]

\paragraph{Term (II).}
Using \eqref{eq:psitilde} again,
\begin{equation}
  (II) = \frac{(p^+)^2}{\sqrt{2\alpha'}}
    \sum_{s\in\mathbb{Z}+\frac{1}{2}}
    \alpha^k_{-\frac12-s}\,\psi^k_s\,\psi^i_{\frac12}\,\psi^j_{-\frac12}|0\rangle.
\end{equation}
\[
\qquad
\begin{cases}
s=\frac12
&
  \alpha^k_{-\frac12-\frac12}\,\psi^k_\frac12\,\psi^i_{\frac12}\,\psi^j_{-\frac12}|0\rangle
\\[0.8ex]&
\alpha^k_{-1}\,\psi^k_\frac12\,\psi^i_{\frac12}\,\psi^j_{-\frac12}|0\rangle
=\alpha^k_{-1}\,\psi^k_\frac12\,[\eta^{ij} - \psi_{-\frac12}^j\psi_{\frac12}^i]|0\rangle=0
\\[0.8ex]
&

\\[1.5ex]
&
\\[2.2ex]

s>0,\ s\neq\frac12:
&
\eta^{ij}\delta_{s,\frac12}=0,
\quad\Rightarrow\quad
\alpha_{-1}^i\psi_{-\frac12}^j\psi_\frac12^i|0\rangle=0
\quad(\psi_\frac12^i|0\rangle=0).
\\[0.8ex]
&

\\[1.5ex]
&
\\[2.2ex]
s=-\tfrac12:
&
\alpha_{-\frac12+\frac12}^i
\Big[
\eta^{ij}\delta_{\frac12,\frac12}
-
\psi_{-\frac12}^j\psi_{\frac12}^i
\Big]\ket{0}
=
\alpha_0^i
\Big[
\eta^{ij}-\psi_{-\frac12}^j\psi_{\frac12}^i
\Big]\ket{0}
\\[0.8ex]
&
=\alpha_0^i\eta^{ij}\ket{0}
-
\alpha_0^i\psi_{-\frac12}^j\psi_{\frac12}^i\ket{0}
\\[0.8ex]
&
=\alpha_0^i\ket{0}
-
\alpha_0^i
\Big(
\eta^{ij}\delta_{0,0}
-
\psi_{\frac12}^i\psi_{-\frac12}^j
\Big)\ket{0}
=0
\\[2ex]
s<-\tfrac12:
&
-\alpha_{-b}^i\,\psi_{-\frac12}^j\,\psi_s^i\ket{0}
=0
\end{cases}
\]

\[
\Rightarrow (II)=0
\]

%\paragraph{Conclusion (NS sector).}
Combining the two results we then get:
\begin{equation}
  \boxed{T^i|j\rangle_{\rm NS} \;=\;
    \bigl(T^i_B + T^i_F\bigr)|j\rangle_{\rm NS} \;=\; 0}
  \label{eq:NSresult}
\end{equation}
Therefore $\Pi^i|j\rangle_{\rm NS} = 0$ in the massless frame, and the
effective little group acting on the massless NS vector is $SO(8)$.

%=========
\section{R Sector: $T^i$ on the Ramond Ground State}
\label{sec:R}
%% =====================================================================

The Ramond ground state $|A\rangle_R$ is massless ($a_R = 0$) and satisfies
\eqref{eq:Rvac}. 

\subsection{Bosonic contribution $T_B^i$}

We now evaluate the bosonic oscillator contribution to $T^i$ on the
Ramond ground state, we have

\begin{equation}
T_B^i|A\rangle_R = i\sum_{n=1}^{\infty}\frac{1}{n} \left[ \underbrace{\alpha^i_{-n}p^+\alpha^-_n|A\rangle_R}_{(I)} - \underbrace{p^+\alpha^-_{-n}\alpha^i_n|A\rangle_R}_{(II)} \right]
\end{equation}

In the Ramond sector and up to the overall normalization convention for $\alpha^-_n$
\begin{equation}
p^+\alpha^-_n
=
\frac{1}{2}
\sum_{m=-\infty}^{\infty}
\alpha^k_{n-m}\alpha^k_m
+
\frac{1}{2}
\sum_{r\in\mathbb{Z}}
\left(r-\frac{n}{2}\right)
\psi^k_{n-r}\psi^k_r 
\label{eq:alpha-minus-R}
\end{equation}

Term $(I)$: 

\begin{equation}
    p^+\alpha^-_n|A\rangle_R= \frac{1}{2}
\sum_{m=-\infty}^{\infty}
\alpha^k_{n-m}\alpha^k_m|A\rangle_R
+
\frac{1}{2}
\sum_{r\in\mathbb{Z}}
\left(r-\frac{n}{2}\right)
\psi^k_{n-r}\psi^k_r|A\rangle_R  \text{ for }  n>0
\end{equation}

Following the same logic as the NS counterpart, there is always at least one positive mode between the two bosonic oscillators and one between the two fermionic oscillators, so $(I)=0$\\

Term $(II)$: the positive bosonic mode annihilates the vaccum directly
%\begin{equation}
$\alpha^i_n |A\rangle_R=0$,
 $n>0$.
%\end{equation}
Hence
\begin{equation}
T_B^i |A\rangle_R=0.
\end{equation}
Thus the bosonic oscillator contribution to the null-rotation generator
annihilates the Ramond massless ground state.

\subsection{Fermionic contribution $T^i_F$}

In the R sector the sum in \eqref{eq:TiF} runs over positive integers
$r \geq 1$. We split into terms (I) and (II):

\begin{equation}
  T^i_F|A\rangle_R
  = i \sum_{r>0} \Bigl[
      \underbrace{\psi^i_{-r}\,p^+\tilde\psi^-_r\,|A\rangle_R}_{(I)}
      \;-\;
      \underbrace{p^+\tilde\psi^-_{-r}\,\psi^i_r\,|A\rangle_R}_{(II)}
    \Bigr].
\end{equation}

\paragraph{Term (I).}
We evaluate $\tilde\psi^-_r|A\rangle_R$ for $r>0$ using \eqref{eq:psitilde}
with $s\in\mathbb{Z}$ and recalling that \begin{equation}   \acomm{\psi^i_r}{\psi^j_s}   =   \delta^{ij}\,\delta_{r+s,0},
\end{equation}
\begin{equation}
  \tilde\psi^-_r|A\rangle_R
  = \frac{1}{\sqrt{2\alpha'}\,p^+}\sum_{s\in\mathbb{Z}}
    \alpha^k_{r-s}\,\psi^k_s\,|A\rangle_R.
\end{equation}
For each $s$:
\begin{itemize}
  \item $s>0$: $\psi^k_s|A\rangle_R = 0$ directly by \eqref{eq:Rvac}.
  \item $s=0$: the term is $\psi^k_r \psi^k_0|A\rangle_R$; since $r>0$,
        $\psi^k_r$ annihilates any ground-state spinor
  \item $s<0$ (write $s=-m$, $m>0$): then $r-s = r+m > 0$, so
        $\alpha^k_{r+m}|A\rangle_R = 0$.
\end{itemize}

Every term vanishes:
\begin{equation}
  \tilde\psi^-_r|A\rangle_R = 0 \;\;\Longrightarrow\;\; (I) = 0.
\end{equation}

\paragraph{Term (II).}
For any $r>0$, $\psi^i_r$ is an annihilation operator on the Ramond vacuum by
definition \eqref{eq:Rvac}:
\begin{equation}
  \psi^i_r|A\rangle_R = 0 \;\;\Longrightarrow\;\; (II) = 0
  \end{equation}

%\paragraph{Conclusion (R sector).}
Combining the two together we obtain:
\begin{equation}
  \boxed{T^i|A\rangle_R \;=\;
    \bigl(T^i_B + T^i_F\bigr)|A\rangle_R \;=\; 0.}
  \label{eq:Rresult}
\end{equation}
Therefore $\Pi^i|A\rangle_R = 0$: null rotations act trivially on the
Ramond ground state, and the effective little group is again $SO(8)$.

%% =====================================================================

\section{Extension to Closed Strings}
\label{sec:Extension to closed Strings}
Our results were done for open strings but can easily be extended to the closed strings case since the Hilbert space of a closed superstring can be viewed as a tensor product of two independent copies of open superstring-like sectors (the left-moving and right-moving sectors). 

For both IIA and IIB superstrings, the full Hilbert space can be written as $\mathcal{H}=\mathcal H_L\otimes\mathcal H_R$.
%\cite{Sen:1986}. 
Consequently, the oscillator contribution to the light-cone generators is additive between the two sectors. Leaving us with,
\[
T^i_{\rm closed}
=
T^i_L\otimes\mathbf 1
+
\mathbf 1\otimes T^i_R .
\]
Thus, for $|\Psi\rangle=|\psi\rangle_L\otimes|\tilde\psi\rangle_R$,
\[
T^i_{\rm closed}|\Psi\rangle
=
(T^i_L|\psi\rangle_L)\otimes|\tilde\psi\rangle_R
+
|\psi\rangle_L\otimes(T^i_R|\tilde\psi\rangle_R).
\]
Since each sector has the same oscillator structure as the aforementioned open-string sector, the previously established result $T^i|\psi\rangle=0$ applies independently to each factor, and therefore
\[
\boxed{T^i_{\rm closed}|\Psi\rangle=0}
\] as well.

The above result also extends to the heterotic theories.  We use the
convention in which the left-moving sector is bosonic, including the
internal gauge CFT, while the right-moving sector is the NSR
superstring.  Thus
%\begin{equation}
$\mathcal{H}_{\rm het}
=
\mathcal{H}^{\rm bos+int}_L
\otimes
\mathcal{H}^{\rm NSR}_R $.
The massless states are of the form
\begin{align}
\alpha^j_{-1,L}|0\rangle_L
 \otimes \psi^k_{-1/2,R}|0\rangle_R,
&\qquad
J^a_{-1,L}|0\rangle_L
 \otimes \psi^k_{-1/2,R}|0\rangle_R ,
\label{hetNS}
\\
\alpha^j_{-1,L}|0\rangle_L
 \otimes |A\rangle_R,
&\qquad
J^a_{-1,L}|0\rangle_L
 \otimes |A\rangle_R ,
\label{hetR}
\end{align}
where $J^a_n$ are the modes of the level-one
$E_8\times E_8$ or $\mathrm{Spin}(32)/\mathbb{Z}_2$ current algebra
\cite{Gross:1985fr}.

As for the IIA, IIB cases, the oscillator part of the Lorentz
generator is additive between the two chiral sectors.  In the frame
$p^j=0$, 
\begin{equation}
T^i_{\rm het}
=
T^i_L\otimes\mathbf{1}
+
\mathbf{1}\otimes T^i_R .
\end{equation}
The right-moving contribution is precisely the NSR operator considered
above, and hence
\begin{equation}
T^i_R\psi^k_{-1/2,R}|0\rangle_R=0,
\qquad
T^i_R|A\rangle_R=0 .
\end{equation}
For the left-moving spacetime excitation,
$T^i_L\alpha^j_{-1,L}|0\rangle_L=0$  follows from the corresponding
bosonic string result \cite{Font:2013}. Only 
the internal state remains to be considered $J^a_{-1,L}|0\rangle_L$.

In the massless frame the left-moving operator takes the form
\begin{equation}
T^i_L
=
i\sum_{n=1}^{\infty}\frac{1}{n}
\left[
\alpha^i_{-n,L}p^+\alpha^-_{n,L}
-
p^+\alpha^-_{-n,L}\alpha^i_{n,L}
\right],
\end{equation}
where the heterotic Virasoro constraint may be written as
%\begin{equation}
$p^+\alpha^-_{n,L}
=
L^\perp_{n,L}+L^{\rm int}_{n,L}$.
%\end{equation}
Therefore
\begin{equation}
T^i_LJ^a_{-1,L}|0\rangle_L
=
i\sum_{n>0}\frac{1}{n}
\left[
\underbrace{
\alpha^i_{-n,L}p^+\alpha^-_{n,L}
J^a_{-1,L}|0\rangle_L}_{(I)}
-
\underbrace{
p^+\alpha^-_{-n,L}\alpha^i_{n,L}
J^a_{-1,L}|0\rangle_L}_{(II)}
\right].
\end{equation}

Term $(II)$ vanishes since
$[\alpha^i_n,J^a_m]=0$ and
$\alpha^i_{n,L}|0\rangle_L=0$ for $n>0$.
For term $(I)$, the transverse contribution to the Virasoro
constraint also annihilates the state, while the internal contribution
satisfies
\begin{align}
L^{\rm int}_{n,L}J^a_{-1,L}|0\rangle_L
&=
[L^{\rm int}_{n,L},J^a_{-1,L}]|0\rangle_L
\nonumber\\
&=
J^a_{n-1,L}|0\rangle_L
=0,
\qquad n>0,
\end{align}
where we used
$[L_m,J^a_n]=-nJ^a_{m+n}$, together with
$J^a_{m>0}|0\rangle_L=0$ and
$J^a_0|0\rangle_L=0$.  Hence
\begin{equation}
T^i_LJ^a_{-1,L}|0\rangle_L=0.
\end{equation}

It follows that both factors in Eqs.~(\ref{hetNS})--(\ref{hetR}) are
annihilated by their corresponding contribution to $T^i$.  Explicitly,
for $|\Psi\rangle=|\phi\rangle_L\otimes|\chi\rangle_R$,
\begin{equation}
T^i_{\rm het}|\Psi\rangle
=
(T^i_L|\phi\rangle_L)\otimes|\chi\rangle_R
+
|\phi\rangle_L\otimes(T^i_R|\chi\rangle_R)
=0 .
\end{equation}
Thus $\Pi^i$ acts trivially on the complete massless spectrum of both
perturbative heterotic theories, and continuous-spin representations
are absent also for
$E_8\times E_8$ and
$\mathrm{Spin}(32)/\mathbb{Z}_2$ heterotic strings.
%% =====================================================================

\section{Conclusions}
\label{sec:conclusions}
%% =====================================================================

We have carried out an explicit, mode-by-mode computation showing that the
null-rotation generators $\Pi^i \equiv M^{-i}$ of the $ISO(8)$ little
group act trivially on all massless states of the NSR superstring in
$d=10$:
\begin{itemize}
  \item In the {NS sector}, $T^i = T^i_B + T^i_F$ annihilates
        $\psi^j_{-1/2}|0\rangle$. The bosonic part commutes past the fermionic
        creation operator and kills the vacuum; the fermionic part
        vanishes because $\tilde\psi^-_r$ acting on the one-fermion state
        always involves an annihilation operator (either a bosonic or fermionic oscillator) acting directly on the vacuum.

  \item In the {R sector}, $T^i$ annihilates $|A\rangle_R$. The
        bosonic part is trivially zero; the fermionic part vanishes because
        every term of $\tilde\psi^-_r|A\rangle_R$ reduces to a
        positive-mode fermion acting on the Ramond vacuum.

        \item The absence of CSPs was also extended to the closed superstrings (IIA, IIB and  heterotic).
\end{itemize}
As such, we have an explicit supersymmetric  extension of the bosonic string result \cite{Font:2013}, that
massless continuous spin representations are absent from the perturbative
superstring spectrum and effectively, the little group for massless superstring
states is $SO(8) \subset ISO(8)$.

 %String theory has thus provided the first-principles explanation of why the physically
%realised little group for massless particles in ten dimensions is $SO(d-2)$
%rather than $ISO(d-2)$.

We consider this to be a straightforward but remarkable result. 
Although string theory is fundamentally a theory of the ultraviolet, it imposes powerful infrared constraints by forbidding an infinite spectrum of massless particles, thereby offering a compelling theoretical explanation for why such states are absent in nature, an experimental result that has no other fundamental explanation.

It is worth emphasizing that this is a distinctly stringy property. The same creation and annihilation operators that build the massless states also generate the entire tower of massive string excitations, each level organized into finite-dimensional representations of the massive little group $SO(9)$; the explicit cancellation between bosonic and fermionic oscillator contributions established in Sections 4–6 is what fixes $T^i=0$, and hence $SO(8)$ rather than $ISO(8)$, specifically at the massless level. This is not a property shared by other higher-dimensional theories: in Kaluza–Klein compactification, for instance, the massless spectrum arises from group-theoretic branching of the parent theory's little-group representations, and so simply inherits whatever helicity or continuous-spin content was assumed for the original higher-dimensional fields, rather than dynamically constraining it. String theory's worldsheet oscillator algebra, by contrast, is fixed independently of any such choice and enforces the absence of CSPs as a dynamical consequence of its structure.

Since our result is intrinsically perturbative, it leaves open the question of whether CSPs play a role in non-perturbative formulations of string theory. M-theory offers a natural setting in which to pose this question: its low-energy limit, eleven-dimensional supergravity, is governed by the couplings among just three massless fields (the graviton, a three-form gauge potential, and the gravitino) and these couplings are well understood perturbatively. We may speculate, however, that a fully non-perturbative formulation of M-theory could require continuous-spin degrees of freedom as well.

There are several possible extensions to this work, naturally we may want to verify the
result explicitly for  non-perturbative string constructions such as string field theory. We leave this to future work.

\paragraph{Acknowledgments.}
We thank Anamar\'{\i}a  Font, Philip Schuster, Stefan Theisen and Natalia Toro for very useful feedback and general discussions. FQ thanks the theoretical physics group at NYUAD for general discussions and advice on the proper use of AI. We counted with the assistance of Claude for checking our computations and for advice on improving the presentation. The research of FQ has been funded by Tamkeen under grant ADHPG-
AD457 to NYUAD.
%% =====================================================================

%\appendix
%\section{Mode Algebra Summary}
%\label{app:modes}
%% =====================================================================

%For completeness we collect the (anti)commutation relations used in
%the main text. For bosonic modes,
%\begin{equation}
  %\comm{\alpha^i_m}{\alpha^j_{-n}} = m\delta^{ij}\delta_{m,n}, %\qquad
  %\comm{x^i_0}{p^j} = i\delta^{ij}, \qquad \comm{x^-_0}{p^+} = -i.
%\end{equation}
%For fermionic modes (both sectors),
%\begin{equation}
  %\acomm{\psi^i_r}{\psi^j_s} = \delta^{ij}\delta_{r+s,0}.
%\end{equation}
%In the NS sector $r,s\in\mathbb{Z}+\half$; in the R sector
%$r,s\in\mathbb{Z}$, so zero modes $\psi^i_0$ with
%$\acomm{\psi^i_0}{\psi^j_0} = \delta^{ij}$ exist.

%% =====================================================================
%% REFERENCES
%% =====================================================================

\end{document}